\documentclass[aps,superscriptaddress,twocolumn,floats,showpacs,prl,amsmath,amssymb,floatfix,nofootinbib,balancelastpage]{revtex4}

\usepackage{graphicx}
\usepackage{subfigure}

\input epsf
\usepackage{psfig}% Include figure files

\begin{document}

\title{Large-Scale Bulk Motions Complicate the Hubble Diagram}
\author{Asantha Cooray}
\affiliation{Center for Cosmology, Department of Physics and Astronomy, 
  University of California, Irvine, CA 92697}
\author{Robert R. Caldwell}
\affiliation{ Department of Physics and Astronomy, Dartmouth College, 6127 Wilder 
Laboratory, Hanover, NH 03755}

\begin{abstract}

We investigate the extent to which correlated distortions of the luminosity distance-redshift
relation due to large-scale bulk flows limit the precision with which cosmological parameters can
be measured. In particular, peculiar velocities of type 1a supernovae at low redshifts, $z < 0.2$,
may prevent a sufficient calibration of the Hubble diagram necessary to measure the dark energy
equation of state to better than 10\%, and diminish the resolution of the equation of state
time-derivative projected for planned surveys. We consider similar distortions of the
angular-diameter distance, as well as the Hubble constant. We show that the measurement of
correlations in the large-scale bulk flow at low redshifts using these distance indicators may be
possible with a cumulative signal-to-noise ratio of order 7 in a survey of 300 type 1a supernovae
spread over 20,000 square degrees.

\end{abstract}
\maketitle

%*************************************************************************************************%

{\it Introduction.}---The challenge to discover the nature of dark energy is pushing all methods
and measures of cosmology to their limits. The luminosity distances to type 1a supernovae (SNe)
which first revealed the cosmic acceleration \cite{Schmidt:1998ys,Riess:1998cb,Perlmutter:1998np},
are now being pursued to obtain tighter constraints on cosmological model parameters
\cite{Tonry:2003zg,Knop:2003iy,Riess:2004nr,Astier:2005qq}. Observational programs, such as the
Supernova Legacy Survey (http://www.cfht.hawaii.edu/SNLS/), the Supernova Factory
(http://snfactory.lbl.gov),  Essence (http://www.ctio.noao.edu/~wsne/), the Carnegie Supernova
Project (http://csp1.lco.cl/$\sim$cspuser1/CSP.html), in addition to ongoing efforts by existing
groups, are currently underway, hoping to achieve $\sim 10\%$ constraints on the dark energy
equation of state parameter. In order to decisively advance our understanding, and test for a
possible time-evolution of the dark energy, a dedicated space-based mission is planned as part of
the NASA/DOE Joint Dark Energy Mission (JDEM).

The luminosity distance-redshift relation, however, has a basic limitation as a tool for cosmology
in an inhomogeneous universe. Large scale structures distort the distances and redshifts. It is
well known that peculiar velocities of SNe induced by the internal properties of host galaxies and
clusters contribute a random component to distance estimates which can be reduced by averaging over
many SNe. Furthermore, gravitational lensing of SN light reduces the accuracy with which the true
luminosity distance can be determined for an individual SN \cite{Gunn:1967,Frieman,HWa,HWb},
thereby complicating an easy interpretation of the Hubble diagram. The effect comes from the slight
modification of the observed SN flux due to lensing by the intervening large-scale structure
\cite{Wang:2002qc,Holz:2004xx,Wang:2004ax,Gunnarsson:2005qu} and correlates distance errors of SNe
spread over the sky at a few degrees or less, due to survey geometries in the form of ``pencil
beams'' or long, but narrow strips \cite{Cooray:05}. Our primary concern in this paper is
large-scale bulk flows \cite{Strauss:1995fz}, peculiar motions that are coherent on scales above
$\sim 60$~Mpc, which correlate individual SN distance estimates spread over ten or more degrees
angular scale. In this case the effect comes from the slight Doppler shifting of both the source
and observer, affecting both the inferred redshift and the flux, resulting in a non-linear
correction to the luminosity distance. This correlated noise cannot be reduced simply by increasing
the sample size and is expected to affect the error budget from low to intermediate  redshifts ($z
< 0.2$). Because the Hubble diagram at these low redshifts must be pinned down accurately in order
that we may hope to find a possible time variation in the dark energy equation of state
\cite{Kim:2003mq,Huterer:2004rf}, it follows that accounting for bulk motions is a necessity.

Fluctuations and anisotropies in luminosity distance have been studied previously, with most of
the focus on formalism \cite{Sugiura,Durrer} and the role of gravitational lensing (e.g.
\cite{Dodelson:2005zt}). The attention has only recently expanded to include peculiar motions
\cite{Hui}. Our intention is to examine the consequences of correlated distortions of luminosity
distances due to bulk motions for the interpretation of the Hubble diagram and efforts to extract
cosmological information about dark energy. Turning the problem around, we will also examine
whether low-redshift SNe can provide a way to measure large-scale bulk flows.

%*************************************************************************************************%
 
\begin{figure*}[t]
\hspace{-0.2in}
\subfigure{\label{fig1a} \includegraphics[scale=0.33]{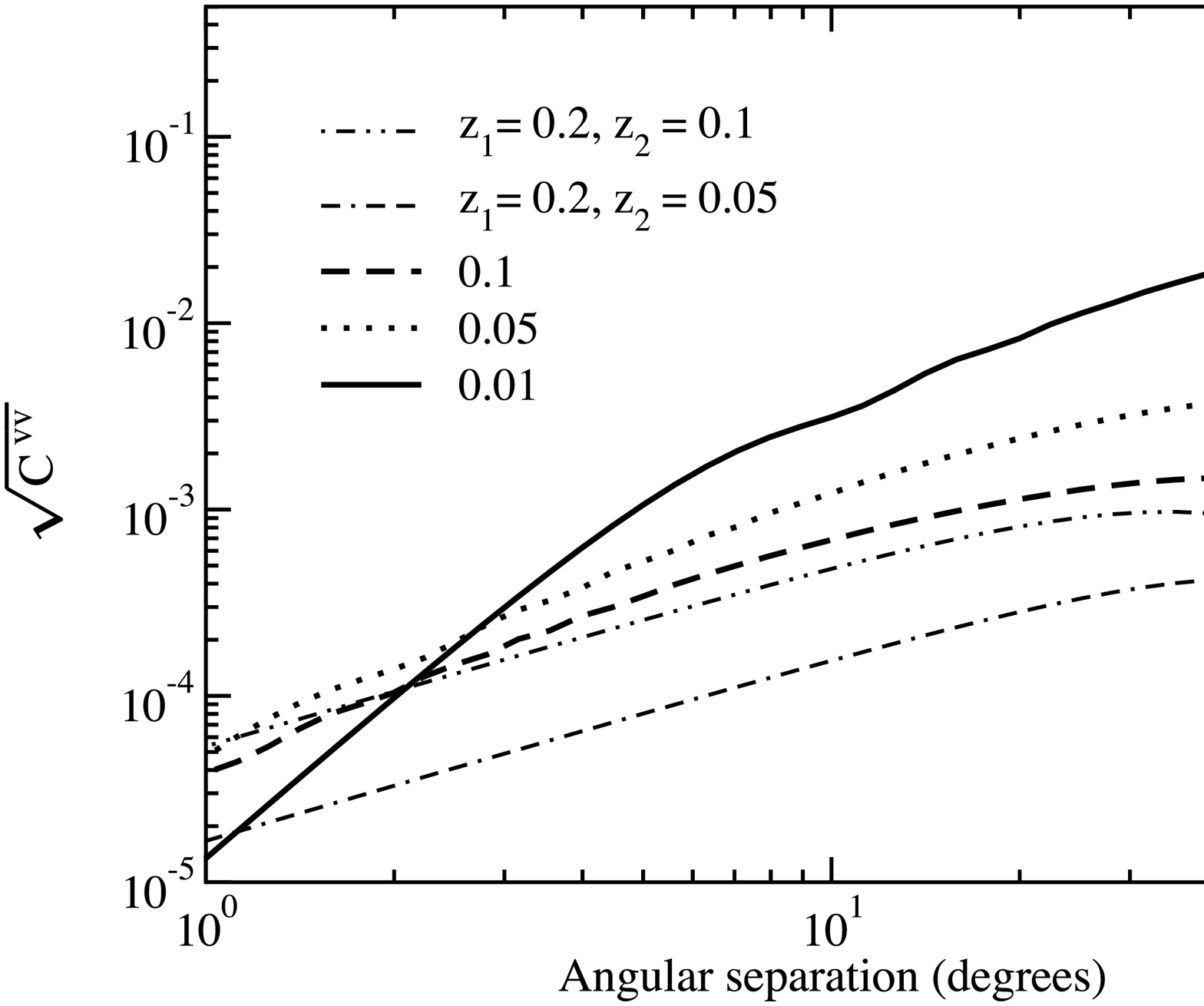}}
\hspace{0.2in}
\subfigure{\label{fig1b} \includegraphics[scale=0.33]{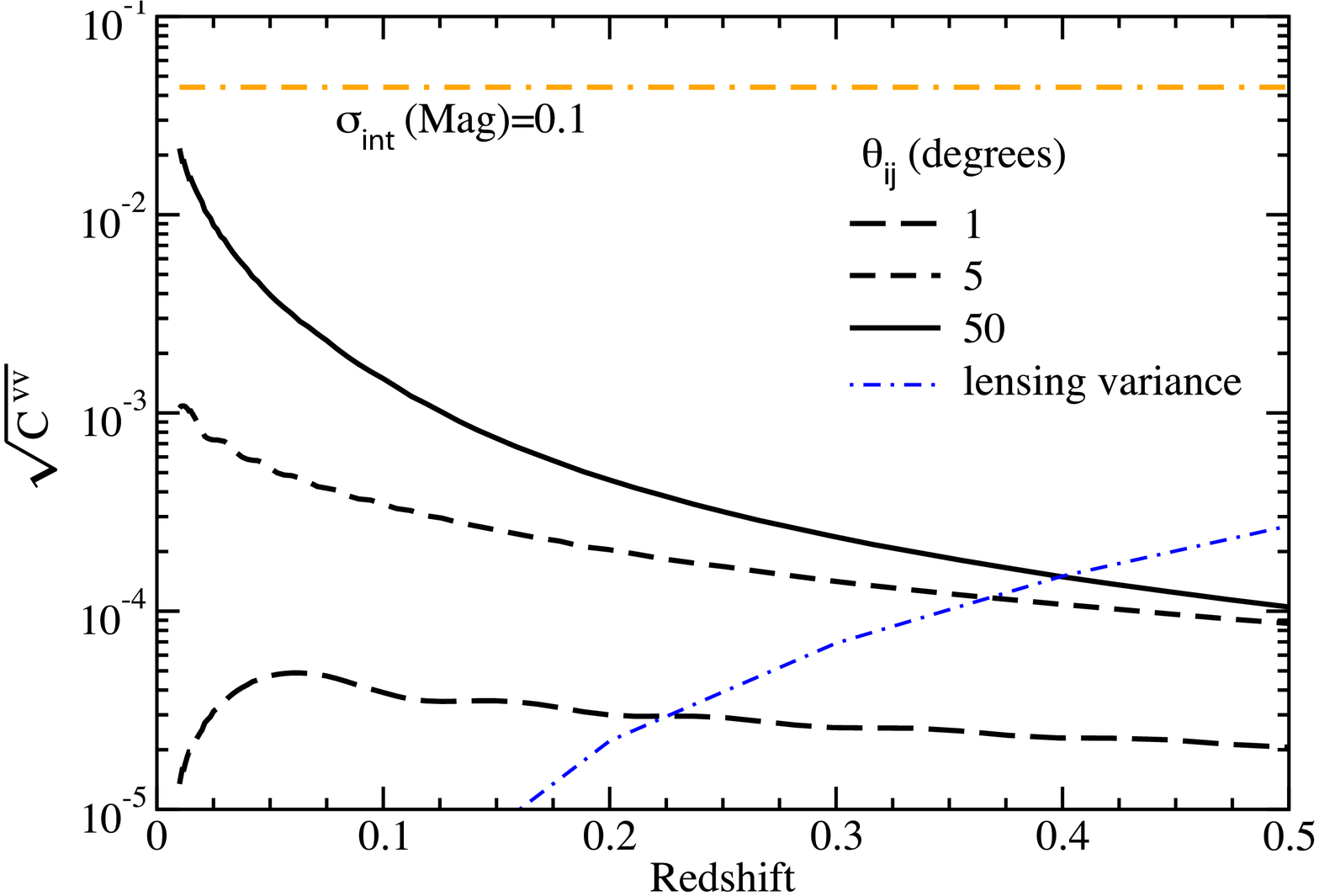}}
\caption{Correlations in peculiar velocities, $C^{\rm vv}(z_1,z_2,\theta)$, as a function of two
SN redshifts, $z_1$ and $z_2$ and their projected angular separation $\theta$. On the left, in
panel (a), we consider the correlations as a function of $\theta$ when $z=z_1=z_2$ and also with
$z_1$ fixed at 0.2 with $z_2$ varied. On the right, in panel (b) we consider the correlations as a
function of redshift, with $z_1=z_2$, for variety of illustrative $\theta$ values. The horizontal
line shows the intrinsic SN measurement error corresponding to $\delta m = 0.1$. Peculiar
velocities correlate SNe separated at 10 to 100 square degrees on the sky at redshifts around 0.1,
but extra covariance from SNe at redshifts greater than a few tenths is negligible compared to the
intrinsic error. The lensing variance overtakes the effect due to correlated motions at redshifts
starting at $z \sim 0.2$, depending on the velocity-velocity separation angle.}
\label{covar}
\end{figure*}

{\it Calculational Method.}---In order to further quantify these statements, we will first
summarize the errors induced by peculiar velocity fluctuations. The effect resulting from
velocities involve two differences: first, the inferred redshift is modified by the difference in
the velocity of the source relative to the observer, projected along the line of sight; second,
the motion at the observer leads to a dipole correction to the distance. In combination, we obtain
(see, Ref.~\cite{Sugiura} for details including their equation 3.15; also \cite{Durrer,Hui}):
\begin{eqnarray}
\frac{\delta d_L}{d_L} = \frac{{\hat {\bf n}}}{c} \cdot \left[{\bf v_{\rm SNe}} - \frac{a}{a' \chi} \left({\bf v_{\rm SNe}} - {\bf v_{\rm obs}}\right)\right] \, ,
\end{eqnarray}
where ${\hat {\bf n}} $ is the unit vector along the line of sight, ${\bf v_{\rm SNe}}$ is the SN
velocity, ${\bf v_{\rm obs}}$  is the velocity of the observer, $\chi$ is the comoving  radial
distance to the SN, and the prime denotes the derivative with respect to the conformal time. The
covariance matrix of errors in luminosity distance is
\begin{equation}
{\rm Cov}_{ij} \approx  \sigma_{\rm int}^2 \delta_{ij} +
C^{vv}(z_i,z_j,\theta_{ij}) \, ,
\label{eq:covar}
\end{equation}
where $\sigma_{\rm int}^2$ is the variance term that affects each distance individually ({\it
e.g.} due to random velocities, or the intrinsic uncertainty in the calibration of SN light
curves). $C^{vv}(z_i,z_j,\theta_{ij})$ is defined as the correlation, at redshifts $z_i$ and $z_j$
with a projected angular separation of $\theta_{ij}$ on the sky, due to velocity fluctuations.
Eq.~(\ref{eq:covar}) defines the full covariance matrix due to peculiar velocities. The covariance
in luminosity distances can be computed, following Refs.~\cite{Gorski,Verde}, whereby
\begin{eqnarray}
&&C^{vv}(z_i,z_j,\theta_{ij}) = \sum_{\rm even \; \; \ell} \frac{2l+1}{4\pi} \cos \theta_{ij}
\frac{2}{\pi} F_l \\
&\times& \int k^2 dk P_{vv}(k,z_i,z_j) 
j_l(k[\chi_i - \chi_j \cos \theta_{ij}]) 
j_l(k \chi_j \sin \theta_{ij}) \nonumber \\
&\times&\left(1-\frac{a}{a'\chi}\right)_i
\left(1-\frac{a}{a'\chi}\right)_j \nonumber
\label{eq:ctheta}
\end{eqnarray}
with $F_l = (l-1)!!/[2^{l/2}(l/2)!] \cos l \pi/2$, and the summation is over even values of $l$.
We assume that SNe are point sources that trace the linear velocity field, but if there is a
velocity bias, then the correlations could be enhanced. Note that $P_{vv}(k,z_i,z_j)$ is the power
spectrum of velocity fluctuations between redshifts $z_i$ and $z_j$ respectively, which can be
written as
\begin{equation}
P_{vv}(k,z_i,z_j) = {D}'(z_i) {D}'(z_j) P_{mm}(k)/k^2
\end{equation}
where $P_{mm}$ is the mass fluctuation power spectrum and $D$ is the mass growth factor. This form
only accounts for linear fluctuations at large scales. The variance related to velocity
fluctuations can be obtained in the limit where $z_i=z_j$ and $\theta_{ij}\rightarrow 0$. We
additionally included nonlinear velocities, corresponding to internal motions of SNe within halos
such as groups and clusters, and found that these also do not affect error estimates. This is due
to the fact that  the velocity-induced variance is smaller than the intrinsic error, $\sigma_{\rm
int}^2$. The effect on the Hubble diagram, however, is not negligible since correlations between
errors  are dominated by the large-scale bulk flows at low-redshifts.

In Figure~1(a), we show the luminosity distance covariance $C^{vv}(z_i,z_j,\theta_{ij})$ with
equal redshifts and also with $z_i=0.2$ for different values of $z_j$, as a function of the
separation angle $\theta_{ij}$. In Figure~1(b) we show the covariance as a function of $z=z_i=z_j$
for several illustrative values of $\theta_{ij}$. For reference, we also plot the variance as a
function of redshift $z$ and compare it to an intrinsic SN magnitude error of $\delta m=0.1$,
which is the expected level to which SN light curves may be calibrated in upcoming searches. We
note that the recent Supernova Legacy Survey (SNLS) has reached an average intrinsic error of 0.12
(in magnitudes). Peculiar velocities are a concern for SNe separated by angular scales of tens or
more square degrees as seen in Figure~1(a), and at low redshifts, $z \lesssim 0.2$, as seen in
Figure~1(b). 

\begin{figure*}[t]
\subfigure{\label{fig2a} \includegraphics[scale=0.33]{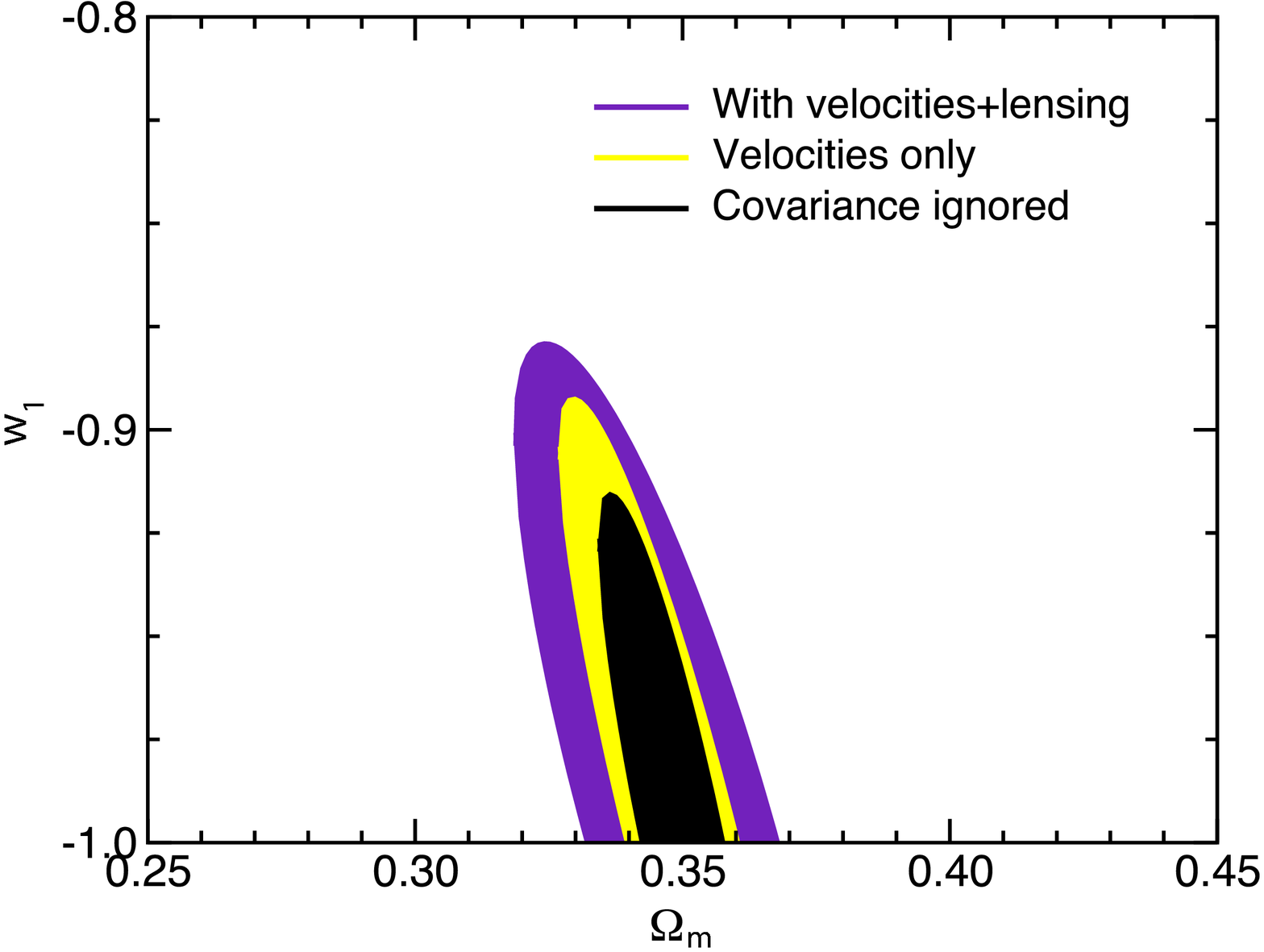}}
\subfigure{\label{fig2b} \includegraphics[scale=0.33]{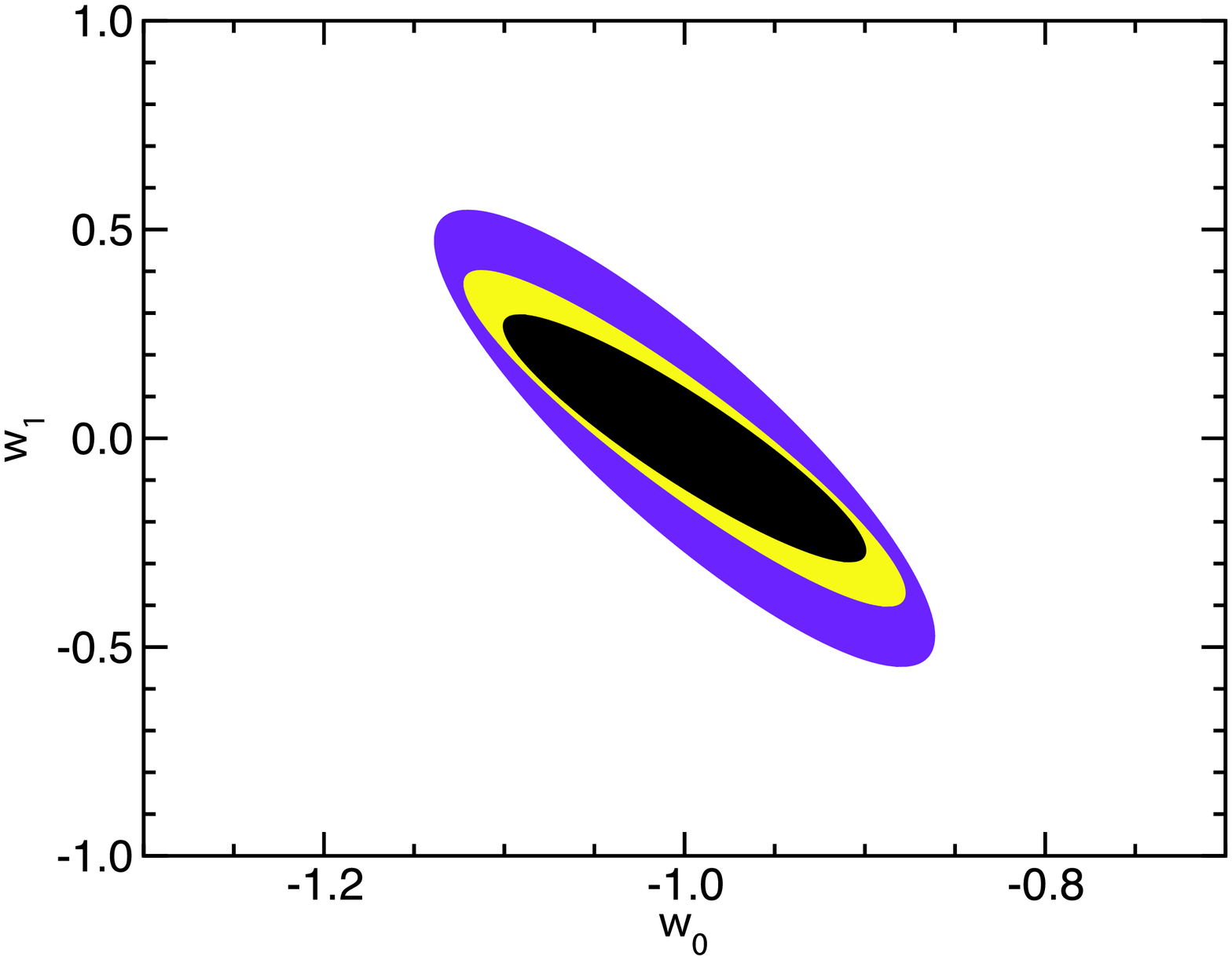}}
\caption{Expected errors on cosmological parameters due to the peculiar velocity
covariances. In Figure~2(a) (left panel) the expected errors on a constant dark energy equation of
state, $w=w_0$, and the matter density parameter are shown. In Figure~2(b) (right panel) the
expected errors on the dark energy equation of state parameters, $w_0$ and $w_1$, where $w = w_0 +
(1-a)w_1$, and assuming a prior uncertainty on $\Omega_M$ of 0.01, are shown. We have assumed a
survey of 300 SNe out to a redshift of 0.2 over 10,000 sqr. degrees, such as the Supernova
Factory, and another 1700 SNe  between redshifts 0.2 and 1.7 in 10 sqr. degrees on the sky, such
as from SNAP/JDEM. We break the error ellipses to covariances from lensing and peculiar
velocities.}
\label{errors}
\end{figure*}

To determine the impact on cosmological parameter estimates, we compute the Fisher information
matrix
\begin{equation}
{\bf F}_{\alpha\beta} = \sum_{ij}
        \frac{\partial d_L(z_i)}{\partial p_\alpha} ({\rm Cov}^{-1})_{ij}
\frac{\partial d_L(z_j)}{\partial p_\beta} \, .
\label{eqn:fisher}
\end{equation}
If the errors are uncorrelated in the Hubble diagram, then the final error on a given cosmological
parameter obtained by model fitting is $\sigma_{\rm int}/\sqrt{N(z_i)}$. But in the case that
there are correlations between data points due to bulk flows, the final error is close to 
$\sigma_{\rm int}\sqrt{1+[N(z_i)-1]r^2}/\sqrt{N(z_i)}$ where $r\equiv {\rm Cov}(i,j)/\sqrt{ {\rm
Var}(i) {\rm Var}(j) }$ is the average correlation between data points. The limit $r\to 0$
corresponds to the case of uncorrelated errors, but in the limit of perfect correlation, $r \to
1$, the error remains as $\sigma_{\rm int}$ with no improvement from the number of SNe in the
survey. For $0 < r^2 < 1$, while there is an improvement with increasing the SNe sample size,
in the limit of large numbers, the  error on an individual parameter will not improve beyond
$r\sigma_{\rm int}$.

%*************************************************************************************************%

{\it Analysis.}---To estimate cosmological parameter measurement errors,  we consider a survey
with 2000 SNe, similar to the combined Supernova Factory and the Supernova Acceleration Probe
(SNAP) proposal for a JDEM. We distribute 300 SNe uniformly in redshift between 0 to 0.2 over an
area of 10,000 sqr. degrees, and 1700 between redshifts 0.2 and 1.7 over 10 sqr. degrees. We
calculate the covariance matrix of size 2000 by 2000 obtained by assuming a distribution of
separations that peaks at roughly one half of the diagonal of the survey geometry (assumed to be a
square).

In Figure~2, we summarize our results related to cosmological parameter estimates. Here, we have
considered the measurement of four parameters, the matter density parameter $\Omega_m$,  the
Hubble parameter $h$ which can also be considered as an overall normalization to the Hubble
diagram (and affected by low-redshift bulk flows),  and assume a dark energy equation of state
given by $w(a)=w_0+(1-a)w_1$. Our fiducial test model is a cosmological constant plus cold dark
matter, with $w_0=-1,\, w_1 = 0$ and matter density $\Omega_m = 0.3$. In Figure~2(a), we assume
$w_1=0$ exactly and consider the measurement of $\Omega_m$ and $w=w_0$. In Figure~2(b), we set a
prior on $\Omega_m$ with $\sigma(\Omega_m)=0.01$, and consider measurement of $w_0$ and $w_1$. The
error ellipses show the expected errors based on which part of the covariance is included. The
innermost ellipse is the case where covariance is ignored and only intrinsic noise is included 
while the outermost ellipse is the case where both peculiar velocities and lensing covariance are
taken into account. We can see that the velocity correlations dilate the $\Omega_m - w_0$
uncertainty by $\sim 25\%$ in the case illustrated by Figure~2(a), and the $w_0-w_1$ uncertainty
by $\sim 20\%$ in the case illustrated by Figure~2(b). Including the effects of both velocities
and weak lensing, for which the variance rather than covariance between sources is dominant, we
see that the uncertainties expand by $40\%,\, 150\%$ on $w_0,\,\Omega_m$ respectively in case (a),
and $50\%$ on $w_0,\,w_1$ in case (b).

In general, smaller separations at high redshift lead to an increase in parameter errors from
lensing, while at low redshifts correlations at the scales of a few tens sqr. degrees increase the
peculiar velocity contribution. A combination of large area ($\sim$ 10,000 sqr. degrees) at $z <
0.2$ and a smaller area ($\sim$ a few tens sqr. degrees) at higher redshifts provides the optimal
combination, though covariances are not simply reduced to zero in that case.

\begin{figure}[t]
\includegraphics[width=3.0in]{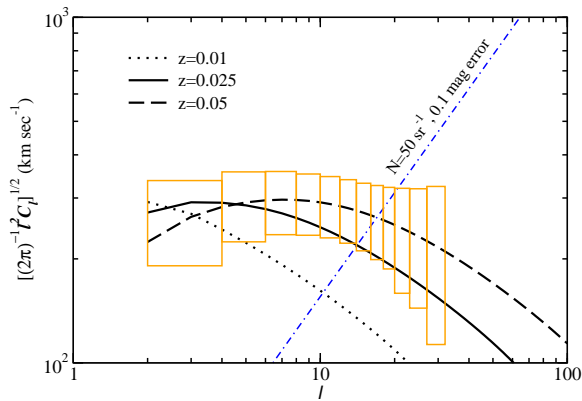}
\caption{The rms fluctuations in the line-of-sight projected velocities $\sqrt{l^2 C_l/2\pi} c$ in
km/s as a function of the multipole. We show the angular power spectrum at different redshifts. The
diagonal, dot-dashed line is the expected noise power spectrum for velocity measurements for a
sample of SNe with an intrinsic uncertainty of 0.1 magnitudes and a surface density of 50 sr$^{-1}$.
Along the lines of a survey such as the Supernova Factory, which expects  $\sim$ 300 SNe at low
redshifts, $z= 0.01 - 0.08$, and assuming the survey to be spread over a fractional sky area $f_{\rm
sky}=0.5$, we plot the expected error boxes for angular power spectrum measurements binned in
multipole space, for the mean survey redshift. The velocity fluctuations are detected at the
cumulative signal-to-noise ratio $\sim$ 7.}
\label{cl}
\end{figure}

There is some possibility to use the dispersion in the Hubble diagram as a measure of peculiar
velocity fluctuations. This is summarized in Figure~3, where we plot the angular power spectrum of
line-of-sight projected velocities as a function of redshift in the form of rms fluctuations given
by $v_{\rm rms}=\sqrt{l^2 C_l/2\pi}c$. The plotted power spectrum is equivalent to the Fourier
transform of Eq.~(3) except that we have not included factors of $(1-a/a'\chi)$ which relate
fluctuations in the velocity field to that of the luminosity distances estimated with SNe.  We are
assuming that an individual distance estimate, combined with redshift and an external estimate of
the overall normalization of the SN light curve, can be converted to an estimate of the peculiar
velocity \cite{Miller:1992}. In Figure~3, we also show the uncertainty related to peculiar velocity
fluctuation measurements, $\sigma_{\rm vel}^2/\bar{N}$ where $\sigma_{\rm vel}$ is the intrinsic
error in the velocity measurement from each SN and $\bar{N}$ is the surface density of SNe (in
sr$^{-1}$).  Assuming an intrinsic  uncertainty of $0.1$ magnitudes, then $\sigma_{\rm vel} =
\sigma_{int}cz/2.17$ (in km/s) which at $z \sim 0.02$ is 275~km/s. We also assume no uncertainty in
the observer's velocity, and that the measurements are not limited by uncertainties in cosmological
parameters such as the Hubble constant or affected by any systematic biases. At low redshifts,
surveys such as the Supernova Factory expect $\sim$ 300 SNe over 2$\pi$~sr so that using an
estimate of 50 sr$^{-1}$ for the surface density and $f_{\rm sky}=0.5$ for the fractional sky
coverage, we obtain the expected error boxes for binned multipole measurements in Fig.~3.  The
line-of-sight projected velocity anisotropy power spectrum is detected with a cumulative
signal-to-noise ratio of $\sim 7$ for the noise curve and error bars shown in Figure~3. However,
this is not a significant detection for detailed cosmological parameter estimates.  For comparison, unlike the low
velocity anisotropy ``signal" captured by low redshift SNe, it may be possible to study clustering
statistics of lensing magnification with samples of SNe at a redshift beyond unity with
signal-to-noise ratios of order thirty or more \cite{Dodelson:2005zt,Cooray:05b}.

While peculiar velocity fluctuations in the Hubble diagram do not provide extra cosmological
information, there are significant implications for distance estimators and cosmological probes.
For example, due to the correlations,  a full $N_{\rm tot}\times N_{\rm tot}$ Fisher matrix, as
opposed to a redshift-binned smaller version, is required in order to obtain cosmological
parameter accuracy estimates. Presumably this is not a problem since the correct treatment of SN
calibration uncertainties already requires the full $N_{\rm tot}\times N_{\rm tot}$ (or even
larger) covariance matrix \cite{Kim_Miquel}. The challenge is significant, however, as one can
neither ignore small correlations nor assume some arbitrary cosmology to estimate covariance among
measurements which are then used to extract new cosmological parameters. The full covariance
matrix must be established as a function of cosmological parameters to obtain an accurate  gauge
of cosmological parameter uncertainty. 

None of these considerations will deter upcoming searches for SNe for cosmological purposes,
though a careful consideration must be given to account for velocity fluctuations at low redshifts
and lensing effects at high redshifts. Since peculiar velocity correlations are only significant
at $z < 0.2$, one can potentially ignore low redshift SNe when fitting distance data to
cosmological estimates. In this case, we find that the parameter errors are not significantly
affected by velocity correlations except that the errors are increased by the fractional factor in
which the SN sample is reduced. In fact, this increase is larger than the case where all
SNe are used to estimate cosmology, but with a proper accounting of the correlations. So, instead
of simply throwing away data, such as low-redshift SNe, it may be best to keep the sample as a
whole, but develop techniques to account for peculiar velocity correlations.

Since the low, $\ell \leq 6$ multipoles in the velocity anisotropy spectrum dominate the SN
distance covariance, if such multipoles can be determined independently of the SN measurements
then corrections, at least partially, can be applied to the interpretations of the Hubble diagram.
If a signal-to-noise ratio of 10 measurement in each multipole is adequate for a reasonable
correction, then independent bulk flow measurements at redshifts ranging over 0.01 to 0.1 must
involve a source surface density of $10^3$~sr${}^{-1}$ and an uncertainty in the velocity
measurement of each object below 100~km s$^{-1}$. Such a surface density of sources and a velocity
error may be achievable with cluster studies of the kinetic Sunyaev-Zel\'dovich \cite{SZpaper}
effect with the upcoming Planck surveyor (http://www.rssd.esa.int/index.php?project=PLANCK),
though foregrounds and internal motions within clusters will contaminate bulk flow measurements
and reduce the overall signal-to-noise ratio levels \cite{Holder:2002wc}. Another approach will be
to consider information from an almost all-sky peculiar velocity survey based on low-redshift
galaxy samples. In the past, the IRAS Point Source Redshift Catalog (PSCz) has allowed modeling of
the spherical harmonic moments of the velocity field \cite{Teodoro:2003ur} out to a redshift of
0.02. We encourage the development of techniques to use information from such surveys to correct
the correlations in the low redshift part of the Hubble diagram.

\begin{figure}[!t]
\subfigure{\label{fig4} \includegraphics[scale=0.33]{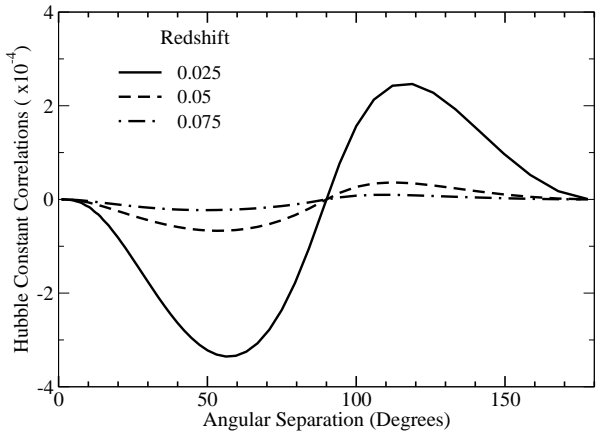}}
\hspace{0.2in}
\caption{Correlations in the Hubble constant, $C^{HH}(z_1,z_2,\theta)$, as a function of $\theta$
when $z_1=z_2$. At redshifts less than 0.05, the correlations are generally at the level of few
times $10^{-4}$ and peak at angular scales of $50^\circ$, suggesting that the Hubble constant
should show fluctuations at the level a few percent, at most, when divided to patches on the sky
at the same angular scale. A reliable detection of this few percent fluctuation is challenging
given the low surface density of SNe expected at low redshifts, similar to the detection of
velocity anisotropies  shown in Figure~3.}
\label{hubble}
\end{figure}

It is interesting to note that the dimensionless fluctuations of the angular-diameter distance
$d_A$ due to large-scale bulk motions are identical to that for the luminosity distance, 
\begin{equation}
\delta d_A/d_A = \delta d_L/d_L .
\end{equation} 
This means our results for velocity covariances apply equally to distances based on
angular-diameter measurements. Possible scenarios include the distances to large scale structure
obtained through baryon acoustic oscillations \cite{Eisenstein:2005su}, distances to galaxies
using the Sunyaev-Zel\'dovich effect (for a recent example, see \cite{Bonamente:2005ct}) or other
features such as radio lobes \cite{Daly:2003iy,Jackson:2003jw}, and probes of cosmology using the
Alcock-Pazcynski \cite{APtest} test which employ the angular-diameter distance to a correlation
radius. The survey details will differ in all cases, so that the extent to which velocity
correlations of low-redshift, wide-separation objects contribute noise will also vary.

We can also consider dimensionless fluctuations of the Hubble constant, inferred from either
luminosity or angular-diameter distances. Using $d_L$, we note that $H^{-1} = d/dz[d_L/(1+z)]$
whereby
\begin{equation}
\frac{\delta H}{H} = -\frac{\delta d_L}{d_L} 
+ \chi \frac{d}{d\eta}\left(\frac{\delta d_L}{d_L}\right). 
\label{dHeqn}
\end{equation}
The second term on the right includes the correction due to the peculiar acceleration. At low
redshifts, this is equivalent to $\delta_H = -\delta_{d_L} + \delta_z (1 + \frac{1}{2}(1-q) z +
{\cal O}(z^2)),$ which we obtain by perturbing the redshift expansion of the luminosity distance.
Here, $\delta_X$ is the fractional perturbation to $X=H,\,d_L,\,z$ and $\delta_z = (1+1/z)\hat{\bf
n}\cdot({\bf v_{\rm SNe}} - {\bf v_{\rm obs}})$. (See Refs.~\cite{Sugiura,Hui}.) In principle, the
deceleration parameter $q$, also varies on the sky and suffers from correlations. Similar to
fluctuations associated with distance in Eq.~(\ref{eq:ctheta}), one can define a covariance for
the Hubble constant anisotropies using the line-of-sight projected correlation function for the
velocity field. This covariance is
\begin{eqnarray}
&&C^{HH}(z_i,z_j,\theta_{ij}) = \sum_{\rm even \; \; \ell} \frac{2l+1}{4\pi} \cos \theta_{ij}
\frac{2}{\pi} F_l \\
&\times& \int dk P_{mm}(k,z_i,z_j) 
j_l(k[\chi_i - \chi_j \cos \theta_{ij}]) 
j_l(k \chi_j \sin \theta_{ij}) \nonumber \\
&\times&\left\{{D}'(z_i)\left(1-\frac{a}{a'\chi}\right)_i + \chi_i \left[
{D}'(z_i)\left(1-\frac{a}{a'\chi}\right)_i\right]'\right\} \nonumber \\
&\times&\left\{{D}'(z_j)\left(1-\frac{a}{a'\chi}\right)_j + \chi_j \left[
{D}'(z_j)\left(1-\frac{a}{a'\chi}\right)_j\right]'\right\} ,\nonumber
\label{eq:htheta}
\end{eqnarray}
where $F_l$ is defined below Eq.~(\ref{eq:ctheta}).  Compared to fluctuations of the luminosity
distance, anisotropies in the Hubble constant are larger by a factor of $\sim 3-5$ depending on
the redshift and the deceleration parameter (see Figure~4).  The increase comes from the
correction to fluctuations associated with peculiar acceleration in Eq.~(\ref{dHeqn}). As shown in
Figure~4, at redshifts between 0.025 to 0.05, fluctuations in the Hubble constant are at most a
few percent, given that the correlations are $C^{HH}(z_i,z_j,\theta_{ij})\sim {\cal O}(10^{-4})$
at angular scales of 60 degrees. Detecting such a small fluctuation from a low redshift SN survey
such as the Supernova Factory, however, will be challenging just as velocity fluctuations are
marginally measurable from SN surveys. This is mostly due to the low surface density of SNe
expected at low redshifts.

Note that the expressions for the Hubble constant and distance fluctuations depend on the
line-of-sight source and observer velocities separately. Hence, there is the possibility of a
``Hubble bubble,'' large fluctuations in $H$, if the motion of the reference frame defined by the
sources, SNe or large scale structure, does not converge to our reference frame, defined relative to
the cosmic microwave background \cite{Davis:1983,Tammann:1990,Sandage:1990,Turner:1992}. For
example, a local low-density bubble could bias $H$ high by $\sim 5\%$ 
\cite{Wang:1997tp,Zehavi:1998gz}, although observations suggest that the reference frames have
indeed converged by length scales $\sim 50$~Mpc/h \cite{Giovanelli:1999,Riess:2005zi}. Nevertheless,
our local motion will induce correlated velocity fluctuations if it is not removed from the data
properly.

To conclude, we have investigated the correlated distortions of the luminosity distance-redshift
relation due to large-scale bulk flows and how these correlations limit the precision with which
cosmological parameters can be measured. At low redshifts, peculiar velocities correlate errors of
type 1a SNe and prevent a precise calibration of the Hubble diagram, relative to the scenario
where one arbitrarily assumes no correlations so that the errors decrease by the square-root of
the number of SNe.  The increase in individual error of a distant SN, or the correction to
variance from the velocity field, however, is negligible relative to an expected intrinsic error
of 0.1 to 0.15  magnitudes. These results are consistent with other recent calculations on how
peculiar velocities affect cosmological studies with SNe \cite{Hui}. Turning our arguments 
around, we find that the measurement of large-scale bulk flows at low redshifts using
SN
distance indicators is challenging. At high redshifts, weak gravitational lensing magnification
adds an extra dispersion to the light curve and increases the individual errors of SN
distance estimates. For surveys that are concentrated on smaller areas on the sky, weak lensing
also correlates distance estimates, but the increase in individual variances generally dominates
the error budget. We also considered similar distortions of the angular-diameter distance, as well
as the Hubble constant. 

\begin{acknowledgments}
We thank Caltech for hospitality, where part of this work was completed. R.C. was supported in part
by NSF AST-0349213 at Dartmouth.
\end{acknowledgments}

\end{document}